\documentclass{article}
\usepackage{spconf,amsmath,graphicx, booktabs,color}

\title{Real Additive Margin Softmax for Speaker Verification}
\name{Lantian Li, Ruiqian Nai, Dong Wang*\thanks{
This work was supported by the National Natural Science
Foundation of China under Grants No.61633013 and No.62171250,
and also the Tsinghua-SPD Bank Joint Research Center for Digital Finance Technologies.
Dong Wang is the corresponding author (wangdong99@mails.tsinghua.edu.cn).}}
\address{Center for Speech and Language Technologies, BNRist, Tsinghua University, China}
\begin{document}
%\ninept
%
\maketitle
\begin{abstract}

The additive margin softmax (AM-Softmax) loss has
delivered remarkable performance in speaker verification.
A supposed behavior of AM-Softmax is that
it can shrink within-class variation by putting emphasis on target logits,
which in turn improves margin between target and non-target classes.
In this paper, we conduct a careful analysis
on the behavior of AM-Softmax loss, and show
that this loss does not implement real max-margin training.
Based on this observation, we present a \emph{Real} AM-Softmax loss which
involves a true margin function in the softmax training.
Experiments conducted on VoxCeleb1, SITW and CNCeleb demonstrated that
the corrected AM-Softmax loss consistently outperforms the original one.
The code has been released at \emph{https://gitlab.com/csltstu/sunine}.
\end{abstract}

\section{Introduction}
\label{sec:intro}

%After decades of research, automatic speaker verification has gained remarkable progress~\cite{campbell1997speaker,reynolds2002overview,hansen2015speaker,bai2021speaker}.
%Traditional speaker verification methods are based on statistical models,
%in particular the Gaussian mixture model-universal background model (GMM-UBM)~\cite{Reynolds00}
%and the subspace alternatives including joint factor analysis (JFA)~\cite{Kenny07} and the i-vector model~\cite{dehak2011front}.
%Recently, deep learning methods have demonstrated great success in speaker verification~\cite{bai2021speaker,ehsan14,li2017deep,heigold2016end,snyder2018xvector,rahman2018attention}.

A key concept of modern speaker verification techniques is to \emph{embed} a variable-length speech utterance into a fixed-length dense vector, usually called \emph{speaker embedding}.
By utilizing the strength of deep neural nets (DNNs) in learning task-oriented features, this embedding
process collects speaker-related information and eliminates other nuances, leading to highly discriminative representations for
speaker traits.

Early research borrows the experience from speech recognition and
employs the classification framework that employs softmax as the activation and cross-entropy as the training objective~\cite{snyder2018xvector}.
By this so-called \emph{softmax training}, vectors of utterances from the same speaker are put together and those from different speakers are repulsed from each other.
Theoretically, this objective is optimal for discriminating
the speakers in the \emph{training} set, but the performance on unseen speakers is not guaranteed.
Unfortunately, speaker verification is an open-set problem in nature, and handling novel speakers is a primary request.
%In other words, the predominant task of speaker verification
%models is not formulating a classification model for the \emph{known} speakers, but establishing a metric space that can well represent
%the distance among \emph{all} speakers. The CE objective is clearly not optimal for this task.

A wealth of research has been conducted to tackle this problem.
One direction is to learn a metric space for speakers~\cite{chung2020defence}, by either pair-based
distance~\cite{heigold2016end}, triplet-based distance~\cite{li2016max,li2017deep}, or
group-based distance~\cite{wang2019centroid,wei2020angular,wan2018generalized}.
This \emph{metric learning} offers an elegant solution for the open-set problem.
However, the training requires large amounts
of pairs or triples, which are not easy to design~\cite{chowdhury2020deepvox}.
Moreover, the training signal produced by pairs or triplets is local and weak,
leading to ineffective training~\cite{bredin2017tristounet,bai2021speaker}.

Due to the problems associated with metric learning, many studies stick on softmax training, but try to remedy its potential weakness on unknown speakers.
The first approach is to involve additional regularization to reduce the
within-speaker variation, for example via the central loss~\cite{Cai2018,li2018deep}, ring loss~\cite{liu2019large} or Gaussian constraint~\cite{li2019gaussian}.
Another approach is to employ various normalization methods to enforce the speaker vectors following some desired distributions,
in particular Gaussian.
This can be conducted either globally or class-dependently.
The former is represented by the famous length normalization~\cite{garcia2011analysis} and the latter
is represented by the deep normalization model based on normalization flows~\cite{cai2020deep}.

Perhaps the most prominent approach is the various max-margin softmax training methods.
Max-margin training aims to maximize the discrepancy between target pair distance and non-target pair distance,
and has been extensively used in metric learning, e.g., ~\cite{li2016max,bredin2017tristounet}.
This idea was adopted to improve softmax training, leading to various max-margin softmax loss,
including Additive Margin Softmax (AM-Softmax)~\cite{hajibabaei2018unified,wang2018additive}
and Additive Angular Margin Softmax (AAM-Softmax)~\cite{xiang2019margin}.
The intuition of these max-margin losses is to make intra-class attraction a more tough task than inter-class repulsion,
so that the target class and non-target class are better separated. In the setting of softmax training,
this means that the difference between target logits and non-target logits should be maximized.

Although the effectiveness has been widely demonstrated, we will show in this paper that these
max-margin softmax losses do not implement a true margin on target and non-target logits.
Specifically, margin, according to the canonical definition, is not only the distance of
two classes, but the region where the loss is incurred~\cite{cortes1995support}. In metric learning based on
triplets, this is often formulated by the following form~\cite{li2016max,li2017deep}:

\begin{equation}
\label{eq:margin}
\mathcal{L}_{\text{margin}} = \text{max} (0, d_p - d_n + m)
\end{equation}
\noindent where $d_p$ and $d_n$ represent the distance of positive and negative pairs, respective, and $m$ is the margin.
According to this definition, the loss is incurred only by hard negative pairs, i.e., negative pairs that
are similar to positive pairs. The margin defined in the max-margin losses mentioned above ignores
the $max$ operation, so does not hold such property.

In this study, we propose a simple modification to AM-Softmax to enforce the true max-margin
training. With this change, the non-target logits will not contribute loss if it has been
much less than the target logit, so that the model will save the parameters to handle other harder non-target
logits. We choose AM-Softmax to conduct the analysis because its form is simple and
has obtained good performance in literature and our own experiments, though the analysis can be similarly
conducted on AAM-Softmax and other margin-based softmax losses.

%Our experiments show that the AM-Softmax with real margin constraint leads to better performance, especially on more challenging tasks.

\section{Methods}
\label{sec:method}

In this section, we firstly revisit the AM-Softmax loss and
discuss its properties and potential limitation, which is followed
by the presentation of our Real AM-Softmax.

\subsection{AM-Softmax}

As a classification loss for training the speaker discriminative DNNs,
the additive margin Softmax (AM-Softmax) loss \cite{wang2018additive} can be formulated as:

\begin{small}
\begin{equation}
\label{eq:am0}
  \mathcal{L}_\text{AM-Softmax} = - \frac{1}{N} \sum_{i=1}^N \text{log} \frac{e^{s(\text{cos}(\theta_{y_i,i}) - m )}}{ e^{s(\text{cos}(\theta_{y_i,i}) -m)} + \sum_{j \neq y_i}e^{s(\text{cos}(\theta_{j,i}))}}
\end{equation}
\end{small}

\noindent where $N$ is the batch size, $m$ denotes the additive margin, $s$ is the scaling factor for training stability. $i$ indexes the training samples,
and $y_i$ is the label of the $i$-th sample.
$\theta_{j,i}$  is the angle between the vector of the $i$-th sample $x_i$ and the representative vector of class $j$ denoted by $w_j$, therefore:

\begin{small}
\begin{equation}
\text{cos}(\theta_{j,i}) = \frac{x_i^Tw_j}{||x_i||\ ||w_j||}
\end{equation}
\end{small}

When implemented by DNN, $x_i$ can be read from the activation of the penultimate hidden layer (before the last affine transform)
and the weights linked to the $j$-th output unit can be regarded as $w_j$. Usually both $x_i$ and $w_j$ are constrained
to be unit-length vectors, so that the cosine distance is computed by simply dot product and implemented as matrix multiplication.
In this case, $\text{cos}(\theta_{j,i})$ is just the logit (i.e., activation before softmax)
on the $j$-th output unit aroused by the $i$-th sample.

\subsection{Margin factor does not boost margin}

Intuitively, due to the margin factor $m$, the AM-softmax enlarges the margin between target logits and non-target logits,
hence leading to a more discriminative model.
The reasoning is as follows: in order to get the same loss as the standard softmax, the target logit in AM-Softmax
must get higher superiority compared to non-target logits.
However, a careful analysis shows this is not exactly the case.

We first make a simple reformulation to the AM-softmax loss as follows:

\begin{small}
\begin{eqnarray}
  \mathcal{L}_\text{AM-Softmax} &=&  \frac{1}{N} \sum_{i=1}^N \text{log} \frac{ e^{s(\text{cos}(\theta_{y_i,i}) -m)} + \sum_{j \neq y_i}e^{s(\text{cos}(\theta_{j,i}))}} {e^{s(\text{cos}(\theta_{y_i,i}) - m )}} \nonumber \\
  &=&   \frac{1}{N} \sum_{i=1}^N \text{log} \big\{1 +  \sum_{j \neq y_i} e^{-s(\text{cos}(\theta_{y_i,i}) - \text{cos}(\theta_{j,i}) - m)}\big\} \nonumber \\
  \label{eq:am}
\end{eqnarray}
\end{small}

\noindent The exponential part of the second term is the most interesting, and it shows that to reduce the loss, the model should enlarge the difference between the target logits
and the non-target logits. This will indeed improve the margin; however, this is not the unique property of AM-Softmax, as the standard softmax can be reformulated in the same way and
the margin is maximized without difference.\footnote{Note that setting $s=1$ and $m=0$ in Eq.(\ref{eq:am}), we can recover the standard softmax.}

Particularly, it seems that the margin factor $m$ is \emph{not} really related to the margin between the target logits and non-target logits, and setting a larger $m$ will not
enforce a larger margin, as it is supposed to do. This can be seen clearly by writing the loss in the following form:

\begin{small}
\begin{equation}
  \mathcal{L}_\text{AM-Softmax}  =   \frac{1}{N} \sum_{i=1}^N \text{log} \big\{1 +  e^{sm} \sum_{j \neq y_i}  e^{-s(\text{cos}(\theta_{y_i,i}) - \text{cos}(\theta_{j,i}))}\big\}
\end{equation}
\end{small}
\noindent It shows that the contribution of $m$ may change the shape of the loss landscape, but not its form. In particular, it is not directly related to the margin between the target
and non-target logits, as defined by $\text{cos}(\theta_{y_i,i}) - \text{cos}(\theta_{j,i})$.

To make it more clear, let us divide the training data into
an easy set and a hard set. We shall also set $s=1$ to expose the contribution of the margin factor $m$.\footnote{Note
this will not limit our discussion as $m$ only appears in the factor $e^{sm}$. Since
$s$ and $m$ are two free parameters, we can always define a new $m$ to absorb $s$.}
Our analysis starts from $m=0$, the special case of the standard softmax.
For the samples in the easy set, the target logit dominates the denominator in Eq.(\ref{eq:am0}), and we have:

\begin{small}
\begin{equation}
\frac{e^{(\text{cos}(\theta_{y_i,i}) - m )}}{ e^{(\text{cos}(\theta_{y_i,i}) -m)} + \sum_{j \neq y_i}e^{(\text{cos}(\theta_{j,i}))}} \approx 1
\end{equation}
\end{small}
\noindent This leads to:

\begin{small}
\begin{equation}
\label{eq:app1}
 1 +  e^{m} \sum_{j \neq y_i}  e^{-(\text{cos}(\theta_{y_i,i}) - \text{cos}(\theta_{j,i}))} \approx 1
\end{equation}
\end{small}
\noindent The loss associated with this sample can be approximated by:

\begin{small}
\begin{eqnarray}
 &\text{log} \big\{1 +  e^{m} \sum_{j \neq y_i}  e^{-(\text{cos}(\theta_{y_i,i}) - \text{cos}(\theta_{j,i}))}\big\} \nonumber \\
 \approx & e^{m} \sum_{j \neq y_i}  e^{-(\text{cos}(\theta_{y_i,i}) - \text{cos}(\theta_{j,i}))}
 \label{eq:bb1}
\end{eqnarray}
\end{small}
\noindent This means when $m$ is increased from $0$, the contribution of easy samples will be emphasized.

For samples in the hard set, the target logit is weak, so we have:

\begin{small}
\begin{equation}
\frac{e^{(\text{cos}(\theta_{y_i,i}) - m )}}{ e^{(\text{cos}(\theta_{y_i,i}) -m)} + \sum_{j \neq y_i}e^{s(\text{cos}(\theta_{j,i}))}} \ll 1
\end{equation}
\end{small}
\noindent This means:

\begin{small}
\begin{equation}
 1 +  e^{m} \sum_{j \neq y_i}  e^{-(\text{cos}(\theta_{y_i,i}) - \text{cos}(\theta_{j,i}))} \gg 1
\end{equation}
\end{small}
\noindent Then the loss associated with this sample is approximated as follows:

\begin{small}
\begin{eqnarray}
&\text{log} \big\{1 +  e^{m} \sum_{j \neq y_i}  e^{-(\text{cos}(\theta_{y_i,i}) - \text{cos}(\theta_{j,i}))}\big\} \nonumber \\
\approx& m + \log \sum_{j \neq y_i}  e^{-(\text{cos}(\theta_{y_i,i}) - \text{cos}(\theta_{j,i}))} \label{eq:hard}
\end{eqnarray}
\end{small}
\noindent This form means that setting any $m$ will not change the optimum.

Overall, setting a large $m$ has the effect of boosting the contribution of easy samples,
while keeping hard samples unweighted. Certainly, this is not a good property as
hard samples are always more concerning. Fortunately, the negative effect cannot be
very disastrous. This is because if $m$ is set overlarge,
the approximation  Eq.(\ref{eq:bb1}) for easy samples will be invalid and turns to the form Eq.(\ref{eq:hard}) of hard samples, hence
losing its impact.

%This indicates that this margin loss cannot obtain robust performance on complicated test conditions.
%Besides, the value of $m$ should be carefully tuned to balance the role of easy samples and hard samples.
%In experiments, we also find that $m$ is sensitive and must be carefully tuned to make the training converged.

%This is certainly a good property.
%However, if $m$ is setting over large, hard examples will have the same approximation form as Eq.(\ref{eq:app1})
%due to the weighting factor $e^{m}$, and so will be attenuated as easy samples.
%This means that $m$ should be carefully tuned otherwise the loss of hard
%and easy examples will be treated equally, and its contribution will be nullified.
%All in all, setting a large $m$ cannot boost the margin between target and non-target logits.
%Since whether a sample is easy or hard changes dynamically during the training process,
%setting a suitable $m$ seems complicated. In experiments, we also find that
%$m$ is sensitive and must be carefully tuned to make the training converged.

Interestingly, the scale factor $s$ seems more effective in boosting the margin.
If we set $s > 1$, then the contribution of the non-targets $j$ will be boosted if
$\text{cos}(\theta_{y_i,i}) - \text{cos}(\theta_{j,i})$ is small, i.e., its contribution
will be amplified. This coincides
the idea of hard negative pair mining, widely adopted in metric learning~\cite{bredin2017tristounet}.

\subsection{Real AM-Softmax}

We analyzed the property of AM-Softmax, and found that the margin factor $m$ does not play the role
of maximizing the margin between target and non-target logits.
Although the scale factor $s$ takes the role partly, the \emph{margin} here is not precise. It
refers to \emph{discrimination}, rather than the canonical definition by Vapnik et al.~\cite{cortes1995support}, 
as shown in Eq.(\ref{eq:margin}).
In this paper, we call this `intuitive' definition of margin in AM-Softmax as \emph{trivial margin}, and the canonical definition
of Eq.(\ref{eq:margin}) as \emph{real margin}. The major difference between these two definitions is that the trivial margin
omits the $max$ operation. According to the analysis in the previous section, this omission leads to serious consequence:
in fact it shuns true max-margin training.

We will follow the definition of real margin to modify the AM-Softmax loss, and
call it \emph{Real} AM-Softmax (RAM-Softmax). This is achieved by slightly modifying the loss function of Eq.(\ref{eq:am})
as follows:

\begin{small}
\begin{equation}
\label{eq:arm}
  \mathcal{L}_\text{RAM-Softmax} = \frac{1}{N} \sum_{i=1}^N \text{log} \big\{1 + \sum_{j\neq{y_i}} e^{\text{max}\{0, -s(\text{cos}(\theta_{y_i,i}) - \text{cos}(\theta_{j,i}) - m)\}}\big\}
\end{equation}
\end{small}

According to this modified formulation, if the target logit is larger than non-target logits by more than $m$, the loss will be zero,
otherwise a positive loss will be incurred. This will encourage the model to focus on hard non-target logits (negative speaker class), and
forget easy non-targets that have been well separated. Therefore, the new loss can cannot be reduced by over-training on easy non-target logits; 
attention must be paid on hard non-target logits. This will balance the contribution of all classes, which arguably alleviates the discrepancy 
between softmax training and the open-set nature in speaker verification.  
Moreover, (real) max-margin training enjoys a solid theoretical advantage in model generalization,
as shown by Vapnik~\cite{cortes1995support}.

Interestingly, if we focus on the exponential part of the RAM-Softmax, the loss is quite similar to the triple loss in 
metric learning, except that all the non-target logits are treated as negative pairs. RAM-Softmax therefore can be regarded as 
a graft of softmax training and metric learning.

\begin{table*}[htb!]
  \vspace{-5mm}
  \caption{EER(\%) results on VoxCeleb1 and SITW. }
  \label{tab:test1}
  \centering
  \begin{tabular}{llccccc}
    \toprule[1pt]
      Objective  & Hyperparameters & VoxCeleb1   & VoxCeleb1-H    & VoxCeleb1-E     & SITW.Dev.Core & SITW.Eval.Core  \\
     \midrule[1pt]
      AM-Softmax & m = 0.20, s = 30 & \textbf{1.739}   & 2.895          & 1.724     & \textbf{2.811}    & 3.362 \\
     \midrule[1pt]
      Real AM-Softmax& m = 0.20, s = 30 & 1.872      & 3.068          & 1.883           & 3.466         & 3.718 \\
                 & m = 0.25, s = 30 & 1.819      & 2.914          & 1.781           & 3.350         & 3.554 \\
                 & m = 0.30, s = 30 & 1.755      & \textbf{2.812} & \textbf{1.696}  & 3.003         & 3.417 \\
                 & m = 0.35, s = 30 & 1.808      & 2.888          & 1.747           & 2.849         & \textbf{3.335} \\
     \bottomrule[1pt]
  \end{tabular}
\vspace{-3mm}
\end{table*}

\section{Experiments}
\label{sec:experiments}

We will compare the AM-Softmax and Real AM-Softmax on several speaker verification tasks.
Note that our purpose is not a SOTA performance; instead, we focus on testing if the `correct' margin works
and how it behaves.

\subsection{Data}

Three datasets were used in our experiments: VoxCeleb~\cite{chung2018voxceleb2},
SITW~\cite{mclaren2016speakers} and CNCeleb~\cite{li2020cn}.
More information about these three datasets is presented below.

\emph{VoxCeleb}: A large-scale audiovisual speaker dataset collected by the University of Oxford, UK.
In our experiments, the development set of VoxCeleb2 was used to train the x-vector models,
which contains 5,994 speakers in total and entirely disjoints from the VoxCeleb1 and SITW datasets.
No data augmentation was used.
\emph{VoxCeleb1} was used as the development set to tune the hyperparameters, and
\emph{VoxCeleb1-E} and \emph{VoxCeleb1-H} were used to test the performance.
Note that the pairs of \emph{VoxCeleb1-H} are drawn from identities with the same gender and nationality, hence
harder to verify than those in \emph{VoxCeleb1-E}.

\emph{SITW}: A widely-used evaluation dataset, consisting of 299 speakers.
In our experiments, \emph{Dev.Core} was used for parameter tuning and \emph{Eval.Core} was used for testing.

\emph{CNCeleb}: A large-scale speaker dataset
collected by Tsinghua University\footnote{http://www.openslr.org/82/}.
It contains more than 600k utterances from 3,000 Chinese celebrities,
and the utterances cover 11 different genres, therefore much more challenging than VoxCeleb1 and SITW.
The entire dataset was split into two parts: \emph{CNCeleb.Train}, which involves 2,800 speakers and
was used to train the x-vector models;
\emph{CNCeleb.Eval}, which involves 200 speakers, was used for testing.

\vspace{-1mm}
\subsection{Settings}

We follow the x-vector architecture~\cite{zeinali2019but} to construct the speaker embedding model,
which accepts 80-dimensional Fbanks as input features and adopts the ResNet34 topology for frame-level feature extraction.
Statistical pooling strategy is employed to construct utterance-level representations.
These representations are then transformed by a fully-connected layer to generate logits and are fed to a
softmax layer to generate posterior probabilities over speakers.
Once trained, the 512-dimensional activations of the last fully-connected layer are read out as an x-vector.
The simple cosine distance is used to score the trials in our experiments.

\vspace{-1mm}
\subsection{Results on VoxCeleb1 and SITW}

In the first experiment, we trained the x-vector models using VoxCeleb2, and then test the performance
on VoxCeleb1 and SITW. The results in terms of equal error rate (EER) are shown in Table~\ref{tab:test1}.
Note that the hyperparameters $m$ and $s$ of AM-Softmax have been carefully tuned, so the performance can be regarded as optimized.
To put focus on margin $m$, we chose the same $s$ in Real AM-Softmax.

Firstly, it can be observed that Real AM-Softmax offers better performance than AM-Softmax
in VoxCeleb1-H, VoxCeleb1-E and SITW.Eval.Core trials, when $m$ was chosen
according to the performance on their own development set. This improvement is not very remarkable
but consistent, demonstrating that the real margin is a correct modification. In particular, since
$s$ is optimized for AM-Softmax, the improvement obtained with real margin can be regarded believable.
Note that with Real AM-Softmax, the performance trend on the development set and the evaluation set
perfectly match, which indicates the training is stable and reliable.

%\begin{table*}[htb!]
%  \caption{EER(\%) results on `hard trials' selected from VoxCeleb and SITW with two objective functions. }
%  \label{tab:test-3}
%  \centering
%  \begin{tabular}{llccc}
%    \toprule[1pt]
%      Objective  & Hyperparameters  & VoxCeleb1-H.H & VoxCeleb1-E.H  & SITW.Eval.Core.H  \\
%     \midrule[1pt]
%      AM-Softmax & m = 0.20, s = 30 & 39.794      & 38.970         & 36.082           \\
%     \midrule[1pt]
%      ARM-Softmax& m = 0.20, s = 30 & 40.729      & 40.416         & 40.206           \\
%                 & m = 0.25, s = 30 & 39.899      & 37.814         & 35.052           \\
%                 & m = 0.30, s = 30 & \textbf{39.175}      & 36.861         & 36.082           \\
%                 & m = 0.35, s = 30 & 39.794      & \textbf{36.821}         & \textbf{32.990}           \\
%     \bottomrule[1pt]
%  \end{tabular}
%\end{table*}

\vspace{-1mm}
\subsection{Results on CNCeleb}

The results in Table~\ref{tab:test1} indicate that Real AM-Softmax is more superior under
challenging test conditions. For instance, the performance improvement with Real
AM-Softmax is more significant on VoxCeleb1-H than on VoxCeleb1-E.

To verify this conjecture, we designed an experiment on CNCeleb, the more challenging dataset.
The x-vector models were trained with CNCeleb.Train, and were tested on CNCeleb.Eval.
Again, the parameters of AM-Softmax were optimized, and we chose the same $s$ in Real AM-Softmax.

Table~\ref{tab:test2} shows the EER results with various settings of $m$
for Real AM-Softmax. It can be observed that Real AM-Softmax outperforms AM-Softmax on
this more challenging dataset, and the relative improvement is slightly more significant than on
VoxCeleb1 and SITW.

\begin{table}[htb!]
  \vspace{-2mm}
  \caption{EER(\%) results on CNCeleb. }
  \label{tab:test2}
  \centering
  \begin{tabular}{llc}
    \toprule[1pt]
      Objective  & Hyperparameters & CNCeleb.Eval   \\
     \midrule[1pt]
      AM-Softmax & m = 0.10, s = 30 & 11.450         \\
%                 & m = 0.20, s = 30 & 11.928         \\
     \midrule[1pt]
      Real AM-Softmax& m = 0.10, s = 30 & 11.618         \\
                 & m = 0.15, s = 30 & 11.323         \\
                 & m = 0.20, s = 30 & \textbf{11.049}\\
                 & m = 0.25, s = 30 & 11.422         \\
     \bottomrule[1pt]
  \end{tabular}
  \vspace{-5mm}
\end{table}

\section{Conclusions}
\label{sec:cond}
\vspace{-1mm}

In this paper we analyze the popular AM-Softmax loss and identify its inability
to conduct real max-margin training. Based on this analysis,
we present a real AM-Softmax loss following the canonical max-margin definition.
With this new loss, hard negative samples are supposed to be more attended.
We tested it on VoxCeleb1, SITW
and CNCeleb, and obtained marginal but consistent performance improvement.
These results demonstrate that the modified real margin function is valid.
Further gains are expected by tuning the scale and margin factors
freely. Moreover, we conjecture AAM-Softmax and other margin-based softmax 
losses can be improved in the same way as they both define trivial margin as AM-Softmax.

\newpage


\bibliographystyle{IEEEbib}
\bibliography{refs}

\end{document}